\documentclass[12pt,a4paper]{article}
\usepackage[left=2.3cm,top=2.4cm,right=2.3cm,bottom=2.4cm]{geometry}
\usepackage{ucs}
\usepackage{authblk}
\usepackage{amsmath,amsfonts,bm,amssymb,latexsym,amsthm}
\usepackage{graphicx, graphics}
\usepackage{multirow}
\usepackage{microtype}
\usepackage[none]{hyphenat}
\usepackage{setspace} 
\usepackage{array}
\usepackage{tikz}
\usetikzlibrary{tikzmark}
\usetikzlibrary{decorations.pathreplacing,calc}
\usepackage{amsfonts}
\usepackage{makecell}
\usepackage[bottom]{footmisc}
\usepackage{color}
\usepackage{url}
\usepackage{float}
\usepackage{subfigure}
\usepackage{rotating}
\usepackage{mathtools}
\usepackage[numbers]{natbib}
\usepackage{hyperref}

\usepackage{morefloats}
\usepackage{booktabs}
\usepackage{multirow}
\usepackage[para,online,flushleft]{threeparttable}
\usepackage[justification=justified,font=small]{caption}

\usepackage{bbding}
\usepackage{pifont}

\usepackage{svg}

\usepackage{footnote}
\usepackage[flushleft]{threeparttable}

\newcommand{\dint}{\mathrm{d}}

\title{Compositional Growth Models}
\author[1,2,3]{Jose Moran}
\author[4]{Massimo Riccaboni}

\affil[1]{\small Macrocosm Inc., Brooklyn NY, USA}
\affil[2]{\small Institute for New Economic Thinking at the
Oxford Martin School, University of Oxford, Oxford, United Kingdom}
\affil[3]{\small Complexity Science Hub, Vienna, Austria}
\affil[4]{\small IMT School for Advanced Studies Lucca, Lucca, Italy}

\date{\today}

\begin{document}

\maketitle

\section*{Introduction}

Traditionally, the analysis of market equilibrium is based on single-product firms. As early as the Fifties, Joan Robinson noted that: ``Dropping the fiction of one-commodity firms destroys the simplicity of the analysis [...] but enlarges its scope'' \cite{robinson1953}, but it was not until the late 1970s that an interdisciplinary trend in the literature developed models for firms consisting of multiple, almost independent units. At that time, however, the models for multi-product firms largely neglected strategic interaction in order to maintain tractability. John Sutton vividly described these models as island models: markets are small islands on which only a single firm fits \cite{sutton1997,sutton2001}. The companies grow by conquering the island markets and behaving like local monopolists. The island economies are independent of each other and there is no local competition. Despite their simplicity, the island models provided useful insights for understanding firm growth, market concentration and aggregate dynamics.

A significant advance in this direction was made at the turn of the century following the seminal contribution of Gene Stanley and coworkers \cite{stanley1996}, with compositional models that consider both the intensive and extensive margin of firm growth \cite{klette2004,fu2005}. The intensive margin is a measure of the size of the company in a given market. The simplest model for this case goes back to \cite{gibrat1931}. By extensive margin we mean the number of almost independent markets in which a company operates. The allocation of business opportunities to firms is usually modeled as a Bose-Einstein process \cite{ijiri1977}. For example, if you assume that national markets are almost independent, the size of a multinational firm is the sum of its sales in the national markets in which the firm operates \cite{armenter2014,bernard2018}. Similarly, the growth of diversified companies with a portfolio of products in different markets is subject to almost independent product-specific shocks \cite{hottman2016}. This approach has been generalized and applied at different levels of aggregation of economic systems. At the finer level, there is the number of transactions (extensive margin) and the value of the good sold in a single transaction (intensive margin). In business-to-business networks, the intensive margin can be represented by the number of customers of a company, while the extensive margin is the amount of revenue per customer \cite{bernard2022}. At the macro level, GDP growth shocks are influenced by firm-specific shocks and the diversification argument does not hold in compositional models of the economy \cite{gabaix2011}. Therefore, in compositional models, shocks at the micro level can have a persistent effect on performance at the aggregate level from firms to the national economies (the so-called \textit{granular} hypothesis).

This chapter summarizes the main advances in the literature dealing with compositional growth models and identifies promising future research directions. The compositional models we describe in this chapter are included in a companion GitHub repository.\footnote{\url{https://github.com/jose-moran/firm_growth}.}

\section{Compositional models of growth}

Consider the size of a firm $i$ at time $t$, $S_{i}(t)$, which is defined as follows

\begin{equation}
 S_{i}(t) = \sum_{i=1}^{K_i(t)}x_{ij}(t),
\end{equation}
where $K_i(t)$ is the number of units and $x_{ij}(t)$ is the size of the unit $j$. The number of units within a firm (i.e. the extensive margin) and the size of each unit (i.e. the intensive margin) are independent and evolve in time \textit{a priori}. In a compositional model, the growth of an aggregated entity, as measured by $S_{i}(t)$, depends on the evolution of the number of units $K_i(t)$ and their sizes $x_{ij}(t)$.

We define the logarithmic growth rate of the firm as $g_i(t) = \log S_{i}(t+1) - \log S_{i}(t)$ and the percentage growth as $r_i(t) = \exp(g_i(t))-1$.\footnote{For the sake of simplicity, we neglect the effect of firm-wide and macroeconomic shocks here. The compositional effect must be combined with aggregate effects to get a complete picture.} We can define the logarithmic $g_{ij}$ and percentage $r_{ij}$ growth rates of the sub-units by replacing $S$ with $x$ in the above equations.

In this case, if we assume for simplicity that $K_i(t+1) = K_i(t)$ we can write

\begin{equation}\label{eq:gr_def}
 r_i(t) = \sum_{i=1}^{K_i(t)} \frac{x_{ij}(t)r_{ij}(t)}{S_i(t)}-1,
\end{equation}
where it is clear that the growth rate of the whole firm is a weighted average of the growth rates of its units. If the number of units $K$ fluctuates, then an additional term must be added to account for this change in size. We will first focus on models where the assumption that $K$ is static in the relevant time frame is justified, and later look at models that explain growth by fluctuations in $K$.

If we assume that the growth rates $r_{ij}$ are independent Gaussian random variables, then $r_i$ is a weighted sum of Gaussian variables and therefore also normally distributed, with variance

\begin{equation}\label{eq:sigma_def}
    \sigma_i^2 = \sum_{j=1}^{K_i}\frac{x_{ij}^2\sigma_{ij}^2}{S_i^2},
\end{equation}
where $\sigma_{ij}^2$ is the variance of the unit growth rates.\footnote{The Gaussian assumption for the $r_{ij}$s can be relaxed to requiring only that they have a finite variance if $K_i$ is large: the sum in Eq.\eqref{eq:gr_def} is Gaussian because of the Central Limit Theorem.}  In other words, conditional on the value of $\sigma_i^2$, the growth rate of the firm is Gaussian. 

In practice, most models study the \textit{logarithmic} growth rather than the percentage growth. However, outside of the tails of the distribution one can approximate $r_i\approx g_i$, and it is often found numerically that many of the statistical properties hold for both objects.\footnote{This excludes the cases of entry/exit or extreme events where the firm's size drops to $0$, leading to $r_i=-1$ and $g_i=-\infty$.}

Because $\sigma_i$ is related to the internal structure of a firm, considering the distribution of $r_i$ across all firms means looking at the superposition of Gaussians with different variances and thus different scales. This is the case, for example, if the firms have a different number of units $K$.
This leads to what is known as a scale mixture of Gaussians, and the shape of the distribution depends on the distribution of the scales $\sigma_i$. Specifically, this means that the distribution of $r_i$ reads $P(r_i) = \int \dint\sigma_i~P(\sigma_i)P(r_i|\sigma_i)$, where $P(r_i|\sigma_i)$ is a Gaussian with variance $\sigma_i^2$.

Scale mixture of Gaussians can give distributions with varying characteristics. For example, if the distribution of volatilities is fat-tailed, with $P(\sigma)\underset{\sigma\gg 1}{\sim} \sigma^{-1-\mu}$ for large $\sigma$ and for some $\mu>0$, then this results in a fat-tailed distribution with the same exponent, $P(r)\sim r^{-1-\mu}$. Similarly, if $P(\sigma)\underset{\sigma\ll1}{\sim} \sigma^k$ with $k>0$, then the distribution is generally more peaked than a Gaussian, and can even present a profile that is not smooth at $0$ if $k<2$. This is the case for a choice $P(\sigma) = \sigma e^{-\sigma^2}\mathbf{1}_{\sigma>0}$ , which leads to $P(r) \propto e^{-|r|}$, the Laplace distribution, which is incidentally also more fat-tailed than a Gaussian. 

\paragraph{What is a unit?} Conceptually and when taking this to the data, we are faced with the problem of defining what a unit is \cite{simon1962}. In the literature, alternative definitions have been used, with market-driven (products or sub-markets), production-driven (establishments or plants), assets (like in portfolio theory) or organizational units (divisions or business units). Firm management is incentivised to split the company into divisions that are roughly independent 
and to diversify activities across independent sub-markets (the so-called island model \cite{sutton1997}). This is done in tandem with the constraint that market forces and other exogenous forces not under the control of management may cause some units to become too large. This may cause the growth rates of the units to be correlated with each other.

Given some arbitrary partition of the firm into units, the growth rate of firm $i$  has a contribution of the form $\sum_j x_{ij}r_{ij}(t)$  with a correlation structure that is e.g. $C_{i,jk} = \text{Cov}(x_{ij}r_{ij}(t), x_{ik}r_{ik}(t))$. This matrix may be diagonalised, with eigenvalues we call $\lambda$, so that $\sum_j x_{ij}r_{ij}(t) = \sum_{k} \sqrt{\lambda_{ik}} r'_{ik}(t)$, where the $r_{ik}'$ random variables are now decorrelated. 
Because the management has incentives to split the company into independent partitions, we expect that $C_{i,jk} = \delta_{jk} + \text{small corrections}$, i.e. that the partition makes the units' growth roughly independent, up to corrections \cite{Sutton2002}. This then leads to $\sqrt{\lambda_{ij}} \approx x_{ij}$, and to a firm volatility that does indeed behave as the formula defined above\footnote{This is in line with the empirical findings of~\cite{mungomoran}}. 

Traditionally, the size of the firm $S$ has been considered to be proportional to the number of units $K$: $S_i=\overline{x}K_i$. Given a finite variance of $x_{ij}$, when the number of units is sufficiently large, the variance of a firm's growth tends to zero approximatively as $K^{-1}\propto S^{-1}$ for the law of large numbers. However, this approximation works only if units are about the same size $\overline{x}$. Empirically, high heterogeneity has been observed to be persistent in time and across different scales of economic systems \cite{stanley1996,riccaboni2008}. Furthermore, the variance is empirically found to decrease much slower, as $S^{-2\beta}$ with $2\beta < 1$.  

The hypothesis that the units are of roughly the same size is therefore not true. This is captured by the Herfindahl index of a firm, which measures how concentrated it is across its sub-units \cite{buldyrev2020}. This quantity is defined as $\mathcal{H}_i=\sum_{j=1}^{K_i}\frac{x_{ij}^2}{S_i^2}$, and is equal to $1/K_i$ all the units have the same size, but of order $\approx 1$ if a single unit concentrates all the size. In the case where the growth rates of units have all the same volatility, which we take to be for simplicity $\sigma_{ij}=\sigma$, then this yields a relationship between volatility and the Herfindahl index that is $\sigma_i = \sigma \mathcal{H}_i$ \cite{gabaix2011}.  

This leads to the intuitive concept of the \textit{effective} number of independent units, $K_{\text{eff},i}:= \mathcal{H}_i^{-1}$, and implies that the volatility goes to $0$ as the inverse of this number rather than the actual number of units. 

Thus, there is a clear relationship between the volatility distribution and the distribution of the aggregate growth rate in the context of compositional models, and since there is a clear link between the structure of firms and their volatility, it is obvious that models specifying different firm structures lead to different distributions of growth rates. We will therefore attempt to explain the different hypotheses in compositional models within a unifying Gaussian scale mixture framework and emphasize how they translate into volatility distributions and thus in different propositions for a growth rate distribution.

\paragraph{Gabaix and Wyart Bouchaud Models}

The models introduced by~\cite{gabaix2011} and by~\cite{wyart2003statistical} imagine a setup where firms have a fixed number of units $K$, and these units have sizes distributed according to a stationary distribution. The main difference between the two models is that \cite{gabaix2011} studies these properties at fixed $K$ while~\cite{wyart2003statistical} consider the possibility that the number of units also fluctuates across firms. In this case and from the discussion above, it is clear that the growth volatility of a firm is directly related to its Herfindahl, as $\sigma_i = \sqrt{\mathcal{H}_i}$.

In both cases, the unit size is distributed with a fat-tailed distribution: $P(x_{ij})\sim x^{-1-\mu}$ for large $x$ and with $1<\mu<2$. This means that the average unit size is well defined, but that its second moment is divergent. As a consequence, when considering the Herfindahl index of a firm $\mathcal{H}_i = \sum_{j=1}^{K_i} \left(\frac{x_{ij}}{S_i}\right)^2$, the largest element of the sum $\sum_{j}x_{ij}^2$ scales as $K_i^{\frac{2}{\mu}}$ while the denominator $S_i^2\sim K_i^2 \mathbb{E}[x]^2$. In other words, the typical value of the Herfindahl scales as $\mathcal{H}_{i,\text{typ}} \sim K_i^{2\left(\frac{1-\mu}{\mu}\right)}$, and since $S_i \approx K_i\mathbb{E}[x]$ one has directly that $\mathcal{H}_i \sim S_i^{2\left(\frac{1-\mu}{\mu}\right)}$. 

In fact, for a fixed value of $K_i$ it is possible to compute the entire Herfindahl distribution $P(\mathcal{H}|K)$, as done in~\cite{mosebo}, to see that it is indeed peaked at $\mathcal{H}_i \approx \mathcal{H}_{i,\text{typ}}$ but that it is distributed as a truncated power-law, i.e. $P(\mathcal{H}|K) \sim \mathcal{H}^{-1-\frac{\mu}{2}}$ for $\mathcal{H}_{\text{typ}}\ll \mathcal{H}\ll 1$, and equal to $0$ for $\mathcal{H}>1$. This implies that the volatility at fixed $K$ has a truncated power-law tail with exponent $\mu$.

Thus, the \textit{average} Herfindahl is dominated by rare observations of firms that have their size concentrated in just a handful of sub-units, which therefore have a much larger concentration than the typical Herfindahl index. Altogether, the average Herfindahl has a contribution $\mathbb{E}[\mathcal{H}|K]\sim K^{1-\mu}\gg \mathcal{H}_{\text{typ}}$, a fact that was overlooked by \cite{wyart2003statistical} and \cite{gabaix2011} but picked up by \cite{mosebo}. However, it is still the case that $\mathbb{E}[\sqrt{\mathcal{H}}|K] \sim \sqrt{\mathcal{H_{\text{typ}}}}$. Going back to firm volatility, this means that one must be careful when defining the object one is working with, as $\sqrt{\mathbb{E}[\sigma^2|K]}\neq \mathbb{E}[\sigma|K]$.  

All in all, one can obtain the growth rate distribution by integrating the Gaussian mixture induced by the Herfindahl distribution. In this case, because the Herfindahl, and therefore the volatility, have a distribution that is a truncated power-law, the growth rate distribution conditional on the number of sub-units $K$ also shares the same features, and it is given by a Lévy alpha-stable distribution that is truncated at $g\sim 1$, as shown by Wyart-Bouchaud. 

Retrieving the entire growth-rate distribution is done by integrating over the distribution of sub-units. Wyart-Bouchaud assume that this number $K$ is itself power-law distributed, as $P(K)\sim K^{-1-\alpha}$ with $1<\alpha<\mu<2$. This results in non-trivial size/volatility relations, with $\mathbb{E}[\sigma|S] \propto S^{\frac{\mu-1}{\mu}}$ but with $\sqrt{\mathbb{E}[\sigma^2|S]}\propto S^{\frac{\alpha -\mu}{2}}$. Note that this second form comes from a contribution where $K$ is not proportional to $S$, since with a probability of order $S^{\alpha-\mu}$ a firm in this model can have a large size $S$ but be made up of only a handful of units, leading to $K\sim 1$ and therefore mechanically to $\mathcal{H}\approx 1$.

Because of its structure, the model is \textit{scale invariant}, in the sense that one can aggregate two firms with $K_1$ and $K_2$ units into one larger \textit{super-firm} with $K_1+K_2$ units, and the growth statistics of that \textit{super-firm} remain well described by the model. This also has a consequence that the firm size distribution in the model has a tail $P(S)\sim S^{-1-\alpha}$. Integrating the $K$-conditional growth rate distribution, one obtains a mixture of truncated Lévy alpha-stable distributions, namely a growth rate distribution with a power-law tail $P(g)\sim g^{-1-\mu}$. 

\paragraph{The Generalized Proportional Growth framework by Stanley and coworkers}

These models, summarized in \cite{buldyrev2020}, postulate that the number of elementary units grows according to the so-called Simon process \cite{ijiri1977}. Keeping with the language of firms, the model imagines $1\leq i \leq N$ firms, each of which has $K_{i}(t)$ units at time $t$.

Between times $t\to t+1$, a new unit arrives in the economy. It becomes a new independent firm with probability $b$, or it is added to an existing firm with probability $1-b$. The unit is then added to firm $i$ with a probability $\propto K_{i}(t)$. At long times, if the process starts in a setting where there are no firms at all, the distribution of $K$ converges to a power-law distribution with an exponential cut-off, i.e. $P(K)\propto K^{-\varphi}f(K)$ where $\varphi = 2+ \frac{b}{1-b}$ and $f(K)\underset{K\to\infty}{\longrightarrow} 0$ exponentially \cite{yamasaki2006}.

However, when $b\to 0$ the picture is different, and the distribution at long times is given by an exponential distribution. This is similar to the Wyart-Bouchaud model, but with a unit distribution $P(K) = \lambda e^{-\lambda K}$ for some $\lambda$. 
Thus, this is also a simple scenario for a Gaussian mixture. In the model of \cite{fu2005}, the distribution of unit sizes is assumed to be log-normal as it results from a random Gibrat-like process. It can then be shown that the \textit{logarithmic} growth rate of a firm with $K$ sub-units is given by a Gaussian with a variance $\propto K^{-1/2}$, namely $P(g|K) \propto \exp\left(-\frac{g^2 K}{2\sigma^2}\right)$. 

This therefore also leads to a Gaussian mixture. In this case, the entire growth-rate distribution reads 

\begin{equation}
    \label{eq:stanley_mixture}
    P(g) \propto \int \mathrm{d} K~ e^{-\lambda K}e^{-\frac{Kg^2}{2\sigma^2}},
\end{equation}
which can be solved to find a distribution that satisfies $P(g)\sim e^{-\vert g \vert}$ for $g\ll 1$ and $P(g)\sim g^{-3}$ for $g\gg 1$. For the case $b > 0$ there is no close form solution for the Gaussian mixture, but in \cite{fu2005} it was shown that the resulting distribution is very similar. In a recent contribution \cite{buldyrev2020}, the model was extended in several directions, for example to take into account a stable economy, the contribution of the change in the number of units $K_i$ and multiple levels of aggregation. For the scaling law, the model predicts a crossover with a variance that does not depend on size for small firms to $S^{-1/2}$ in the limit $S\to\infty$ \cite{riccaboni2008}.

This model can be extended to \textit{exponential mixtures of Gaussians}~\cite{Buldyrev2007}. This corresponds to the case where $P(K)=\lambda e^{-\lambda K}$, but where the variance is set to be a general power of $K$, so that $P(g|K) \propto \exp\left(-\frac{g^2}{2\sigma^2 K^\psi}\right)$. The case $\psi=-1$ corresponds to the situation described above, but other distributions can be obtained. 

For instance, the choice $\psi=-2$ would lead to a distribution behaving as $P(g)\sim g^{-2}$ for large $g$, and more generically power-law tails are obtained for $\psi<0$.  The choice $\psi=1$ leads directly to a Laplace distribution, and corresponds to a different firm growth model that is not a compositional model and that was studied by \cite{Bottazzi2005Laplace}. In this model, firms are made up of one single unit, but at each time-step a fixed number of growth opportunities are available to them. These growth opportunities are attributed at random, leading to a Bose-Einstein distribution, which has an exponential tail. Therefore the number of company-wide growth shocks a firm receives per time unit has an approximately exponential distribution, but now the growth volatility scales with the number of growth shocks and leads to the exponent $\psi =1$. This can be understood by saying that the volatility of a company's growth scales directly with the number of growth shocks per unit of time it receives, but inversely with the number of units it has.

\begin{figure}
    \centering
    \includegraphics[width=\textwidth]{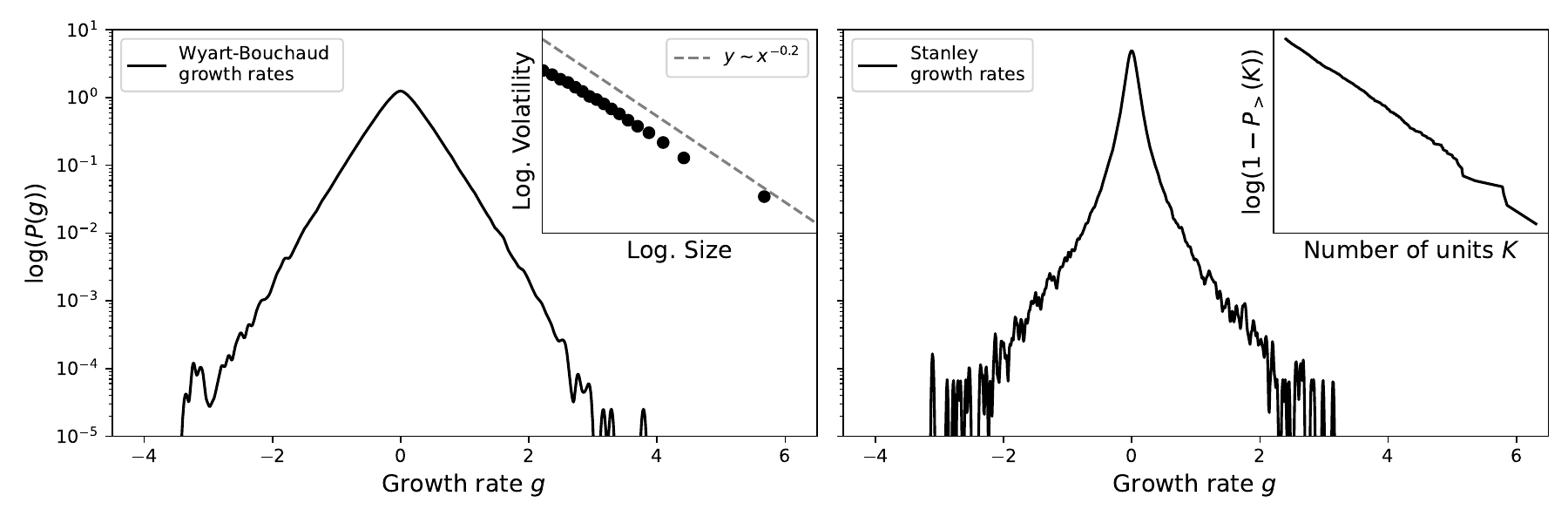}  
    \caption{\textit{Left:} Results for a simulation of the Wyart-Bouchaud model~\cite{wyart2003statistical} with $\alpha=1.2$ and $\mu=1.4$, resulting in $\frac{\alpha-\mu}{2}=0.2$. The main plot shows a Gaussian KDE estimation of the density, showing fatter-than-Gaussian tails, while the inset shows the size volatility relation, which decays to $0$ slower than $S^{-1/2}$. \textit{Right:} Results for a simulation of the Stanley group's model with $b=0$. The main plot shows a Gaussian KDE estimation of the density, with fat tails. The inset shows the distribution of the number of units per firm in the simulation of the model.}
    \label{fig:growth_rates}
\end{figure}

\paragraph{Sutton's model}
The goal of the model put forth by~\cite{Sutton2002} is to explain the anomalous scaling between size and volatility. To this end, he imagines a \textit{microcanonical} model where a firm is taken to be of an integer size $S_i$, and is partitioned into $K$ units with sizes $x_{ij}$ such that $\sum x_{ij}=S_i$. Sutton assumes that all partitions (both in terms of number and in terms of size) are equiprobable, and uses results on partitions of the number of integers to compute his result. The same results are obtained in a very succinct fashion in~\cite{wyart2003statistical}.

The result is that, conditional on firm size, a firm has on average $\exp(a\sqrt{S_i})$ units, with $a>0$, and that their sizes are distributed according to a Bose-Einstein distribution, with $p(x_{ij})\propto \left( \exp\left( \frac{b}{2} x_{ij}/\sqrt{S_i}\right) - 1\right)$.  Because this is again a compositional model, the unit size distribution can be used directly to compute the variance of the percent growth conditional on size. The result is that the variance scales as $\sigma(S)^2\propto S^{-1/2}$, providing an explanation to the anomalous scaling with an exponent $\beta = 1/4$.

Rescaling the growth rate of a firm as $r/\sigma(S)$, one finds that the distribution of this quantity approaches a Gaussian distribution as $S\to\infty$ regardless of the distribution of the growth rates of individual sub-units. However, it can be shown that the excess kurtosis decays very slowly with $S$, so that important corrections to the Gaussian asymptotic case can be seen for finite $S$.

\paragraph{Farmer-Axtell-Schwarzkopf} 
This model, proposed by~\cite{schwarzkopffarmer}, is different to the others in the sense that it explains growth not by varying the size of the sub-units in time, but by varying their number. In this model, all the units have a constant size ) and the equality $S_i(t)= K_i(t)$ holds up to a constant multiplicative factor, which we set to $1$ for simplicity. Thus, the percentage growth reads $r_i(t) = K_i(t+1)/K_i(t) - 1$, and the entire process is explained through the dynamics of $K_i(t)$.

The idea of the process is that each time step, an unit is replaced by $n$ new units drawn at random from some distribution $P(n)$. Thus, for a fixed $K_i(t)$, $K_i(t+1)$ is the sum of a series of draws from the distribution $P(n)$, and $r_i(t) = \frac{1}{K_i(t)}\sum_{j=1}^{K_i(t)} n_{ij}-1$. If one picks a fat-tailed distribution, i.e. $P(n)\sim n^{-1-\mu}$ with $0<\mu\leq 2$ for large $n$, then under the generalised Central Limit Theorem the distribution of $K_i(t)^{-\frac{1-\mu}{\mu}}r_i(t)$ converges to a Lévy alpha-stable distribution with parameter $\alpha$.  

This model is then used to derive the volatility-size scaling relationship. The key idea is that $K_i(t+1)$ is dominated by the largest draw from $p(n)$, which scales as $K_{i}(t)^{1/\mu}$. Taking the square, we get the scaling $\mathbb{E}[K_i(t+1)^2]\sim K_i(t)^{2/\mu}$, which can be reworked easily to yield $\sqrt{\mathbb{V}[r_i(t)^2]}\sim K_i(t)^{\frac{1-\mu}{\mu}}$ and therefore to the size-volatility scaling relation $\sqrt{\mathbb{E}[\sigma^2|S]}\sim S^{\frac{1-\mu}{\mu}}$.

On the surface, this size-volatility scaling relation looks similar to the one obtained for the Wyart-Bouchaud model, although here the model is not a Gaussian mixture. The trajectory of firm growth is wildly different between the two models, size the model proposed by~\cite{schwarzkopffarmer} says that the growth-rate of a single individual firm is a Lévy flight, and therefore implies that from one period to the next the size of a firm can fluctuate by various orders of magnitude, which seems an unrealistic model of a firm. 
In contrast, the Gaussian mixture model such as the ones described before suggests instead that the growth rate of a single firm would follow something closer to a standard geometric Brownian motion, with Gaussian fluctuations that have a firm-dependent volatility. This leads to more tamed and realistic trajectories, and the anomalous statistics observed in firm data are the result of wild fluctuations in firm heterogeneity rather than wild fluctuations in firm trajectories.

\section*{Concluding discussion}

Over the past twenty years, compositional growth models have played a central role in various areas of economic analysis, from industrial organization to international trade, from economic theory to macroeconomics. The scaling properties of economic systems have been fundamentally rethought. A key feature of compositional models is the granular hypothesis: economic systems cannot be broken down into units of approximately the same size. This is a characteristic distinctive feature of complex systems whose constituent parts have a skewed size distribution across all levels of aggregation. Some products are blockbusters, others are fiascos, a few customers are much more important than others, a handful of countries and companies dominate the global economy. At the company level, compositional models help to explain deviations from Gibrat's growth model. At the macro level, these models provide an explanation for the excessive volatility of composite economies whose effective diversification is bounded away from the predictions of the law of large numbers.

Despite recent progress in the analysis of firms consisting of a varying number of almost indipendent units of different sizes there are still some promising research directions to explore. First and foremost, the effects of strategic interaction and constraints on unit growth must be adequately considered. In the current version of compositional models (\cite{klette2004}), the entry of a new product into a market implies that another product from a different company is displaced. However, several products may be on the market at the same time and their market shares may be interdependent (see \cite{sutton2007} for an interesting contribution in this direction). In general, cross-price elasticity and product interdependence must be taken into account. There are also some technological interdependencies in the product space, as emphasised in the economic complexity literature. A natural extension of compositional models is the consideration of more complex dynamics of corporate growth such as mergers and acquisitions (M\&As), see also~\cite{Lera2017} for first steps in that direction. One way to include M\&As is to consider the possibility of companies acquiring more than one business opportunity at a time in order to take over (parts of) already existing companies. Another possible extension is the integration of compositional models and production networks. Currently, compositional models are based on simplyfing assumptions about the production function that firms use to convert inputs into outputs.
Accordingly, with rare exceptions, the change in company size is usually measured on the basis of the number of employees or total turnover. The dual problem of changing the composition of inputs in business-to-business production networks needs to be investigated in future research. More in general, we must consider the evolution of firm sizes in a shrinking economy. Compositional growth models were initially developed for expanding economies with a growing number of individuals, products and firms. More recently, versions of these models have considered stable economies with steady state distributions of firm sizes. However, given the demographic transition, we need to develop models for business dynamics in shrinking economies.

Another important line of research concerns the use of machine learning methods to predict firm growth and aggregate dynamics \cite{bargagli2021supervised}, or the use of firm dynamics to infer network information to be used in economic modelling~\cite{mungomoran}. The increasing availability of fine-grained data sets at the unit level across different industries and countries makes it possible to combine machine learning with compositional models to improve economic forecasting. On a more broad note, macroeconomic forecasting can take advantage of data-driven agent-based methods, and compositional models allow to derive aggregate predictions using ensemble models of different learners at the unit level.

\bibliographystyle{apalike}
\bibliography{bibio}

\end{document}